# Reducing Message Collisions in Sensing-Based Semi-Persistent Scheduling (SPS) by Using Reselection Lookaheads in Cellular V2X

**Yongseok Jeon, Seungho Kuk and Hyogon Kim * 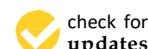**

Department of Computer Science and Engineering, Korea University, Anam-Dong, Sungbuk-Gu, Seoul 02841, Korea; ysjeon741@korea.ac.kr (Y.J.); shkuk@korea.ac.kr (S.K.)
* Correspondence: hyogon@korea.ac.kr; Tel.: +82-2-3290-3204

**Abstract:** In the C-V2X sidelink Mode 4 communication, the sensing-based semi-persistent scheduling (SPS) implements a message collision avoidance algorithm to cope with the undesirable effects of wireless channel congestion. Still, the current standard mechanism produces a high number of packet collisions, which may hinder the high-reliability communications required in future C-V2X applications such as autonomous driving. In this paper, we show that by drastically reducing the uncertainties in the choice of the resource to use for SPS, we can significantly reduce the message collisions in the C-V2X sidelink Mode 4. Specifically, we propose the use of the "lookahead", which contains the next starting resource location in the time-frequency plane. By exchanging the lookahead information piggybacked on the periodic safety message, vehicular user equipment (UEs) can eliminate most message collisions arising from the ignorance of other UEs' internal decisions. Although the proposed scheme would require the inclusion of the lookahead in the control part of the packet, the benefit may outweigh the bandwidth cost, considering the stringent reliability requirement in future C-V2X applications.

## 1. Introduction

The automotive industry and the information technology (IT) industry are joining forces towards connected and autonomous vehicles that offer a multitude of benefits, such as driving safety, traffic flow efficiency, driver comfort, and new infotainment experiences for passengers. Along with sensors and computing intelligence, a key component in the development is the vehicle-to-everything (V2X) communication, which allows a vehicle to communicate with not only other vehicles and road-side equipment, but also nearby pedestrians and the Internet. Among the prospective applications of connected and autonomous vehicles, the most crucial one that strongly drives the V2X technology evolution is the safety. Analyses by the U.S. Department of Transportation's National Highway Traffic Safety Administration (NHTSA) show that V2X communication can address up to 80 percent of the crash scenarios involving non-impaired drivers [1].

In vehicular safety communication, the periodic exchange of Basic Safety Messages (BSMs) containing such information as position, speed, acceleration, and heading among others [2], is indispensable for reducing the collision risk by facilitating the tracking and short-term prediction [2] of neighboring vehicles' kinematics. It immediately enables driving safety applications such as Forward Collision Warning (FCW) [3]. For long, the Dedicated Short-Range Communication (DSRC) based on the IEEE 802.11 Outside-the-Context-of-BSS (OCB) mode [4] has been the sole candidate lower layer technology for the vehicular communication. Both the IEEE WAVE framework in the U.S. [5–7] and the Intelligent Transport Systems (ITS) G5 framework in the Europe [8] build on the DSRC radio.

Recently, however, the V2X communication using the cellular communication infrastructure is emerging as a strong contender of the DSRC-based communication. Although some argue that cellular V2X (C-V2X) is not as ready, has compatibility problems between different 3GPP Releases, and has exaggerated coverage when measured in reality [9], proponents of the C-V2X argue that the cellular alternative has multiple advantages over DSRC [10–13]. Although DSRC is a mature technology and does not require a coordinating network infrastructure or bulky signaling procedures, it has weaknesses in the spectral efficiency [14] and the capability to deal with congestion and hidden terminal problem at scale [15]. Typically, C-V2X is considered superior to DSRC in these aspects: longer range [16,17] and enhanced reliability, more consistent performance under traffic congestions, clear evolution path towards 5G for emerging applications [14], and better coexistence with other technologies. On the other hand, the higher packet reception performance of C-V2X over DSRC is highly affected by the modulation and channel coding scheme (MCS) [18], and the capacity can be lower than DSRC under very high vehicle density and short distances [16]. Therefore, C-V2X must be less reliant on the cellular infrastructure for efficiency, by leveraging the sidelink communication in particular.

Cellular V2X communications have already been standardized by the 3rd Generation Partnership Project (3GPP), in Release 14 [19]. It is focused on the communications for basic safety use cases. In particular, the sidelink (SL) device-to-device (D2D) communications through the PC5 interface are a key enabling technology. There are two modes, Mode 3 and Mode 4, that were designed for direct V2X communication. The difference between the two modes is in the radio resource allocation method. Resources are allocated by the cellular network under mode 3. Mode 4 does not require cellular coverage, and vehicular user equipment (UEs) autonomously select their radio resources using a distributed scheduling scheme with a message collision resolution mechanism. Mode 4 is considered the baseline mode and represents an alternative to DSRC.

C-V2X lacks the contention resolution mechanism like that of the IEEE 802.11 in the DSRC communication. Therefore, if two UEs happen to choose the same time-frequency resource for BSM transmission, collision cannot be avoided. Moreover, it cannot be detected by either party. Worse yet, the sidelink Mode 4 uses Semi-Persistent Scheduling (SPS), the collision episode can persist across multiple messaging intervals without the colliding UEs knowing it. Furthermore, vehicles may choose one of a few standardized messaging intervals (e.g., 10 Hz, 20 Hz, or 50 Hz [20]), raising the possibility of the UEs with the same messaging interval facing the repeated message collision risk. This is one of the reasons that the scheduling is not persistent, but semi-persistent. Each "streak" of periodic message transmissions has an average duration of one second, after which a UE moves to other resource with some probability for the next streak [21]. This resource reselection will part the unknowingly colliding UEs. Another reason for the SPS is the topology change. When new vehicles join the group of vehicles that are coordinating the resource use through the sensing-based SPS, a newly joining vehicle may have been using the same time-frequency resource that another vehicle from the existing group uses. In this case, the same sustained message collisions can ensue.

Although the current standard has the aforementioned message collision resolution mechanism, we observe that it has not been engineered to its maximum potential to meet the message delivery probability for safety communications. Especially when we envision a high-reliability C-V2X communication for future applications such as remote driving and autonomous driving, we need a very tight resource coordination and control among vehicular UEs. In this paper, we show that by sharing the information regarding their resource reselection to neighboring UEs, each UEs can mutually lower the message collision probability. Specifically, we propose the vehicles broadcast in their safety beacons their planned resource reselection earlier than their actual reselection instance. We show through simulation that the proposed algorithm far exceeds the packet delivery ratio (PDR) performance in various situations. We also discuss what changes should be made in the current standard to carry the planned reselection information. In particular, we discuss how the SCI Format 1 should be modified for better message collision resolution.

In this paper, we limit ourselves to the study of the *algorithmic aspect* of the message collision resolution mechanism in the sensing-based SPS algorithm. Other aspects such as the physical channel models and their impacts to the algorithm are deferred to a future work. Also, note that the message collision resolution itself is not a congestion control mechanism, but it is directly related with the congestion control in C-V2X. When congestion becomes severe, message collisions will become more likely, and an efficient collision resolution algorithm will mitigate the undesirable impact of the high congestion level on the packet delivery performance. However, devising a fully-fledged congestion control algorithm using one or more standard-provided means such as power control, MCS control, retransmission control, or packet dropping, is beyond the scope of this paper.

The rest of the paper is organized as follows. In Section 2, we summarize the works that compare C-V2X with the incumbent IEEE 802.11p DSRC/C-ITS. Then we discuss recent related work on radio resource management, network-controlled and autonomous, where our work falls in the second category. In Section 3, we describe the sensing-based SPS algorithm and its collision resolution mechanism employed in the current standard specification. In Section 4, we discuss the proposed scheduling algorithm. In Section 5, we perform simulation experiments to compare the current SPS and the proposed algorithm, showing the performance gap we can obtain through the proposal. In Section 6, we conclude the paper with the summary of our work and the list of future work items.

## 2. Related Work

Within the context of C-V2X, radio resource management is an actively studied topic. Recent works in this topic are classified into network-controlled and autonomous approaches. For the former, Sun et al. [22] transformed the latency and reliability constraints of C-V2X into optimization constraints that are computable using only the slowing varying channel state information. They also formulated a problem to optimize the performance of both vehicular UEs and cellular UEs, and proposed an algorithm to solve it. Abanto-Leon et al. [23] considered the resource allocation problem under Mode 3. When there is conflict in the allocation between two overlapping groups of vehicles as in an intersection or merging highways, some vehicles cannot receive each other's information. They designed algorithms with different complexity and performance to solve the problem. Zhang et al. [24] considered the problem of reusing the same resource block to maximize the number of concurrent V2V communications. Under eNodeB's control, the proposed algorithm can improve the spectrum efficiency, so that LTE can support a high density of vehicles. Cecchini et al. [25] considered a network-controlled resource management based on the vehicle position information. They exploited the concept of minimum reuse distance, at which the same resource can be used by a different transmitter without affecting those receivers that are in the awareness range. They showed that the accuracy of the localization affects the error rate. Fritzsche and Festag [26] also proposed location-based scheduling, where the base station scheduler can leverage the optimal scheduling distance to maximize the cell throughput via resource reuse among different vehicles. To achieve the objective, they exploited the relations among cell throughput, reliability, and communication range. Kim et al. [27] proposed a position-based resource allocation scheme that allocates a different frequency and time resources based on vehicle speed, density, direction, and position. They showed that the scheme can improve the packet reception ratio (PRR). Şahin et al. [28] proposed how to allocate resource for the out-of-coverage vehicles when they enter such area. They assumed that the network infrastructure makes the prediction as to the movements of the vehicles, and accordingly makes the reservation for them. They showed that the prediction accuracy is important for transmission success.

As vehicles can be in locations where there is no infrastructure support, autonoumous resource allocation is an indispensable mechanism. Consequently, works are emerging that attempt to evaluate or improve on the 3GPP C-V2X Mode 4. Bazzi et al. [29] analyzed through simulation the impact of parameters on its wireless resource allocation performance. They varied five physical-layer (PHY) and three medium-access layer (MAC) parameters and classified them into three groups: those having negligible impact, those affecting the quality of service, and those creating the trade-off relation. In particular, they found that the resource keeping probability creates the trade-off between the PRR and update delay. Wang et al. [30] analyzed the collision probability and average delay in the SPS under the assumption of perfect PHY performance. They found that a flexible resource block selection enables a trade-off between the delay and the collision probability. Specifically, they found that by letting the UEs choose the closest resource block, the delay in the reselection event can be significantly reduced at the cost of the moderately increased collision probability. He et al. [31] showed that by separating the control and the corresponding data packets in the time axis and by letting them carry the reservation information for each other, the collisions are significantly reduced. This is the most closely related work with ours, but we notice that to carry the resource reservation in the data packet, a cross-layer cooperation is required between the radio layer and the safety application layer. Our work proposes an enhancement within the currently standardized sensing-based SPS framework. Bonjorn et al. is another closely related work with ours. It evaluated the performance of the SPS as proposed in the 3GPP specification [32]. To lower the collision probability, the authors proposed a cooperative resource allocation and scheduling solution. In the proposal, each UE informs the others of its resource reselection counter (RC) value and resolve any expected collision by forcibly changing the RC values of the UEs that have the same RC value. By reducing the probability of overlapping the reselection windows of the involved UEs, the authors showed that the average collision rate is significantly lowered. Moreover, they proposed to eliminate the randomness in the parameters such as the probability of retaining the same resource in the next streak, and the RC counter. An important finding in this work is that most collisions prevented by this scheme are caused by newly incoming vehicles to the communication range. A deficiency in this work, however, is that the considered resource use level of 25% is rather low to be considered congestion. Moreover, always reselecting other resource when the RC counter reaches zero could raise the collision probability when the use level is high as we will show in Section 5.2.2. In this paper, we consider various levels of resource use including the severe congestion situations. Also, we attempt to broadcast more specific information as to the location of the next resource to be used by a UE instead of the RC counter. Molina-Masegosa and Gozalvez [33] presented a detailed analysis of the performance of LTE-V sidelink Mode 4 and proposes a modification to its distributed scheduling. This work showed that under heavy congestion, the IEEE 802.11p at a high rate transmission [34] can outperform LTE-V in terms of the PDR. The work noticed the differences in the BSM messages as one source of such poor performance by LTE-V. As a solution to this problem, the authors suggested that no subchannels are reserved for a streak when transmitting the larger and less frequent packets (e.g., 300 bytes). Nabil et al. analyzed the effect of the Mode 4 resource pool configuration and some of the key SPS parameters on the scheduling performance and found that the resource reservation interval (RRI) significantly influences packet data rate performance, whereas the resource keeping probability has little effect in dense vehicular highway scenarios [35]. The reason is that when the congestion is severe, there remain only a limited amount of available resources. Due to lack of adequate resources, UEs end up choosing the same resources again and this would be equivalent to maintaining the resource reservation without executing the SPS algorithm. The authors showed that that proper configuration of scheduling parameters can significantly improve performance. They concluded that research on congestion control mechanisms is needed to further enhance the SPS performance for many practical use cases. Our work confirms most of their findings, but contrary to this work, we also find that the resource keeping probability parameter has visible effect in a wide range of congestion as we will show in Section 5.2.2.

## 3. Resource Scheduling and Collision Resolution in Sidelink Mode 4

The resource reservation in the sidelink Mode 4 works in a distributed manner among UEs, so it provides a distributed mechanism to mitigate message collisions [19]. Specifically, the 3GPP standards define parameters and possible mechanisms to cope with the channel congestion [20,21], which is called the sensing-based SPS. SPS is prescribed by the standards due to the periodic nature and the predictable sizes of the safety messages such as BSM transmitted by a UE. Under the SPS scheme, therefore, a group of resources evenly spaced by the RRI on the time axis are reserved in one signaling exchange. For instance, a UE can be allowed to transmit ten BSMs each spaced 100 ms before another attempt to reserve resources for the next group of BSMs is made. Typical RRI values are 20 ms, 50 ms, and 100 ms, where longer intervals are also allowed. The other type of resource reservation, namely dynamic scheduling, requires signaling for each packet transmission. It has much larger signaling overhead, thus can delay the transmission of the time-critical safety messages.

To determine which single-subframe resource(s) a UE can use, it relies on *sensing* the resource use by UEs in the recent past. By allocating the resource using the sensing result, the SPS scheme tries to minimize the message collisions as congestion worsens. However, the sensing-based SPS should not be considered a fully-fledged congestion control algorithm, which would use various available means such as power control, MCS control, retransmission control, or packet drop [20,36]. Below, we briefly discuss the sensing-based SPS algorithm as defined in the current standards.

### 3.1. Sensing

The resource scheduling in C-V2X is done over a two-dimensional space, where the two axes are time and frequency (Figure 1). The wireless resource grid is divided into subchannels in frequency and subframes in time. In the frequency axis, the granularity of resource allocation is a subchannel. Each block in the grid (e.g., small colored boxes in the figure) is called the single-subframe resource, and it is composed of multiple resource blocks (RBs). Since there is no central coordinator for the resource scheduling in SL Mode 4, each UE monitors the resource usage by other UEs before selecting the resource for its BSM transmission. This is called *sensing*. The resource that has been used (or is predicted to be used) by other UEs are marked busy and not selected for transmission to prevent the message collision.

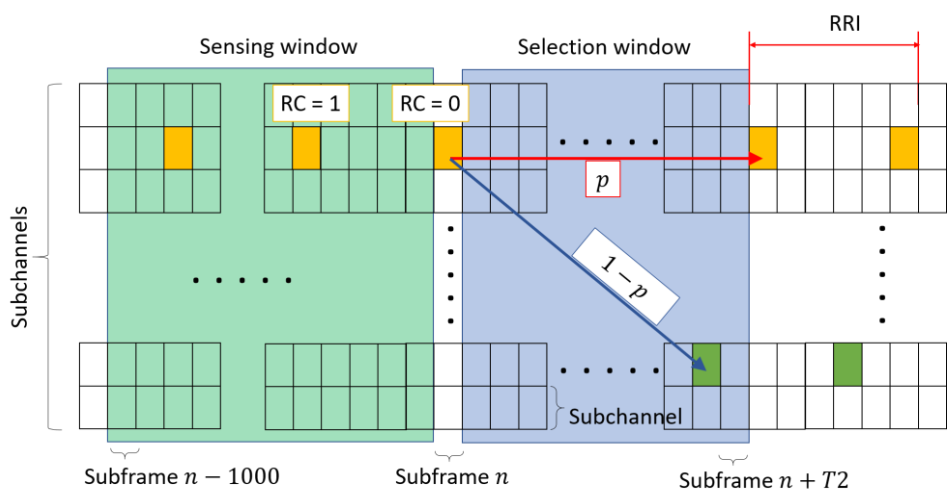

**Figure 1.** Reselection in SPS; $p$ is the resource keeping probability.

In the sensing window, 1000 subframes (1 s) in the immediate past are monitored. For pedestrian devices, *partial sensing* is allowed for battery saving. Whether a given resource is being used is determined based on the RSSI threshold called the sidelink RSSI (S-RSSI) threshold, above which a single-frame resource is considered busy. The standard does not specify the value of this threshold, but the 3GPP working documents usually compute this threshold by adding −107 dBm/RB in the subchannel [37]. Based on the sensing result, the next selection step determines which single-subframe resource it should exclude from the selection window as unusable. Essentially, the resources that have been used by other UEs in the sensing window enable us to determine if a resource in the selection window will likely be available.

### 3.2. Selection

Suppose a UE $V$ is at subframe $n$. For a BSM transmission, $V$ needs to reserve new subchannel(s) whose number depends on the packet size at a subframe in $[n + T1, n + T2]$, where ($T1 \leq 4$) and $T2$ ($20 \leq T2 \leq 100$) [21]. The interval between $n + T1$ and $n + T2$ is called the *selection window* (Figure 1). In this window, candidate single-subframe resources (CSRs) are selected after filtering out two unusable groups of resources. First, for the single-subframe resource $m$ in the selection window, a UE marks it unavailable if $V$ was transmitting in subframes $m - RRI \cdot j$ ($j \geq 1$) in the sensing window. This is because the subframes at which $V$ transmitted its packet could not be sensed by $V$ due to half-duplex transmission. Because there might be other UEs that transmitted in these subframes, the subframe $m$ is excluded to avoid possible collisions with these UEs. Second, we exclude the resources that will probably be used by other UEs. These are the single-subframe resources $m^t$ in the same subchannel(s) where $m^t - RRI \cdot j$ ($j \geq 1$) were used. Whether a resource was used or not is determined by a reference signal received power (RSRP) sensed higher than the threshold corresponding to the priority of $V$'s packet to be transmitted. The priority is provided by the higher layers [21,36]. After excluding these two groups of resources, the remaining resources are called the set $S_A$.

In case the single-subframe resources in $S_A$ is less than 20% of the entire selection window, then more candidate resources should be identified. For this, we raise the RSRP threshold by 3 dB, and repeat the filtering process for the second group. If $S_A$ becomes larger than 20%, then we choose those with the smallest RSSI values, which we call $S_B$. We report $S_B$ to the higher layer that randomly selects one of them for the first transmission. The number of times that the subchannel(s) for the selected resource is used without the selection process is also randomly selected and is called SL_RESELECTION_COUNTER [20], or simply RC counter in this paper.

### 3.3. Reselection

Each packet transmitted in a streak, each RRI apart from the previous one, decrements the RC counter by one. The streak length, namely the RC counter, is randomly set to one value in [$C1, C2$]. The range of the random number depends on the RRI. It is [$C1, C2$] = [5, 15] if RRI is 100 ms or higher, in [10, 30] if RRI is 50 ms, and in [25, 75] if RRI is 20 ms. Notice that the average length of the streak is designed to last only a second, hence semi-persistent scheduling.

When RC reaches 0, the next streak should be scheduled. This operation is called *reselection*. First, a decision is made as to whether a different resource should be chosen from the selection window using sensing again or the current resource location is kept. In the latter case, the same RRI is maintained between the last packet in the current streak and the first packet in the next streak. The decision is controlled by the resource keeping probability *probResourceKeep* that the upper layers configure [20]. For the decision, a random number in the interval [0, 1] is generated, and if it is larger than *probResourceKeep*, a different resource should be used for the next streak. Otherwise, the UE continues to use the current resource. In either case, however, a new RC counter value should be decided as above, according to the RRI. Figure 1 shows an example where the UE is at the point of deciding whether it should change the resource location with probability $(1 - p)$ or it should maintain the current resource for another streak with $p$. The standards stipulate that the probability $p$ in the figure, *probResourceKeep*, should be configured to be between 0 and 0.8.

As we discussed in Section 1, there are two reasons why we keep reselecting resources. With only a few choices of RRI [20], without any collision resolution mechanism similar to DSRC, and with multiple packets transmitted in a single streak, UEs can collide multiple times without knowing it. Therefore, the streak is designed to last only a second on average so that any colliding UEs can part from the other in a second. The packet collisions can also occur when new vehicles join the group of vehicles that are already coordinating the resource use through the sensing-based SPS.

In this paper, we focus on this reselection step to improve the packet collision mitigation performance. Thus, we assume that $S_B$ has been already computed according to the standard. Algorithm 1 describes the reselection part in the SPS algorithm, to be used for our simulation experiments below.

---

**Algorithm 1** Reselection by a UE in SPS

---

1: **procedure** SPS ($RRI, T1, T2, C1, C2, N_{subCH}, probResourceKeep$)
2:      $txSubCH \leftarrow random(1, N_{subCH})$                                *1>* Initializations for $txSubCH, txSubframe, RC$
3:      $txSubframe \leftarrow random(1, RRI)$
4:      $RC \leftarrow random(C1, C2)$
5:      $subframe \leftarrow 0$                                                                     *1>* This is the current time
6:  – – – – – – – – – – – – – – – – – – – – – – – – – – – – – – – – – – – –
7:      **while** *True* **do**
8:          **if** $subframe == txSubframe$ **then**                                               *1>* It is time to transmit
9:              $txPacket(txSubCH)$                        *1>* Transmit packet on the specified subchannel(s)
10:             **if** $RC /= 0$ **then**                                                 *1>* Not time for reselection yet
11:                 $txSubframe \leftarrow txSubframe + RRI$                     *1>* Schedule next packet in one RRI
12:                 $RC \leftarrow RC - 1$
13:             **else**                                                              *1>* Time to reselect for next streak
14:                 $RC \leftarrow random(C1, C2)$
15:                 **if** $random(0, 1) < probResourceKeep$ **then** *1>* Should keep the same resource location
16:                     $txSubframe \leftarrow txSubframe + RRI$        *1>* Maintain the same RRI across streaks
17:                 **else**                                         *1>* Must move to other location for the next streak
18:                     call **select_resource( )**                  *1>* Reselect $txSubCH$ and $txSubframe$; Section 3.2
19:                 **end if**
20:             **end if**
21:         **else**
22:             call **sensing_update( )**                *1>* Keep sensing and update resource use map; Section 3.1
23:         **end if**
24:         $subframe \leftarrow subframe + 1$                                                        *1>* Push time
25:     **end while**
26: **end procedure**

---

In Algorithm 1, $txSubCH$ is the subchannel where the UE will transmit its BSMS in the next streak. In line 2, it is initially randomly selected from among $N_{subCH}$ subchannels in the carrier. In line 18, when the reselection happens, a new resource location will be chosen. Otherwise, the current location will be kept (line 15). In case the resource location must be changed, the sensing-based SPS discussed in Section 3.2 is performed. The subframe index of the BSM transmission, $txSubframe$, is also randomly selected at the beginning. It is incremented by RRI (line 16). The length of the next streak is always randomly selected (lines 4, 14). In line 22, the resource use map is updated using the sensing procedure described in Section 3.1.

## 4. Sensing-Based SPS Scheduling with Lookahead

### 4.1. Motivation

We notice that one reason the current SPS algorithm produces significant packet collisions [33,35] is the *uncertainty* about the next resource location a UE will (re)select. In fact, there is no provision in the current specification of the sensing-based SPS with which a UE can convey the information about its selected location for the next streak in the time-frequency plane. Although randomness is necessary to spread the resource selections thus reduce collisions, there is randomness cost to it—the collision probability cannot be minimized below a certain limit. Although the sensing provides some information about the likely locations to be used in the selection window, it is insufficient information to deterministically prevent a collision. For example, a sensed use of a resource location $(c, m)$ by a UE does not mean that $(c, m + RRI)$ will be occupied, where $C$ is a subchannel and $m$ is a subframe. Because the UE-internal variable RC is not advertised to other nodes ([38], Section 5.4.3.1.2), such newly occurring vacancies cannot be exploited. Worse yet, a resource location in the selection window that a UE $V$ selected when its RC counter reaches 0 at subframe $n$ may be selected by other UEs in $[n - (T2 - T1), n]$. A UE does not have any means to avoid such probabilistically arising collisions, because it does not know other UEs' RC values. It is why Bonjorn et al. [32] proposes to piggyback the internal RC counter value on the broadcast packet, using which the UEs with the simultaneously expiring RCs try to reduce collision probabilities by intentionally changing the RC values and realigning their selection windows. However, realigning the entire selection windows is costly, as significant realignment cannot be made under the latency requirement ($T2$ is set according to the requirement). Compared with the wide window size as large as 100 subframes, the realignment through the change of RC values is limited by the RC range, which can be as small as five (subframes). The realignment may further be blocked by many potentially colliding UEs when the congestion level is high. Even if we can overcome all these hurdles, if selection windows overlap, the resource selection will still be probabilistic.

### 4.2. SPS with Lookahead Scheme

We propose that we take a more direct approach to removing uncertainty in the SPS algorithm while not undermining the necessary randomness in standard parameters. Specifically, we let each packet carry the resource location information for the next streak of packets. Moreover, we require each UE determine the location before the current streak ends. Please note that in the original SPS algorithm, the location is determined just before RC reaches 0 [20]. We call this scheme *SPS/LA* for "SPS with lookahead". Unlike Bonjorn et al. [32], we try to realign only the individual single-subframe resources with the same individual resource location determined as the starting location for the next streak. Since such information is available earlier than actual movement to the planned location, UEs can have chances to change their decision and advertise the newly adjusted location. In this paper, we let UEs determine the next location at $RC_{LA} = 1$. This is because larger $RC_{LA}$ values face more uncertainty in the observation of resource usage. Namely, the resource reservation information by other nodes through their published lookaheads and the current use pattern are subject to changes until $RC = 0$, especially when the congestion level is already high.

The advertised location information for the next streak, the lookahead, is composed of three fields that point to the next resource location and size of the first packet of the next streak. Namely, it is

$$LA = <c, L, n>$$

where $c$ is the starting subchannel, $L$ is the number of consecutive subchannels it uses to transmit the packet (a.k.a. transport block (TB)), and $n$ is the subframe index, of the resource location. Notice that other UEs can easily infer the current RC value of the advertising UE, because the last packet in the current streak will be in $[n - (T2 - T1 + 1), n - 1]$.

Figure 2 shows an example where the lookahead helps avoid the message collision that would be unavoidable in the original SPS. In Figure 2a, UEs $A$ and $B$ collides at a resource in subframe $n^t$ since they do not know each other's internal decision as to the location of the next streak. In contrast, in Figure 2b, $B$'s decision to use the resource is conveyed in the lookahead, which $A$ notices. It changes its earlier decision to move to the contended resource, and it moves to a different resource location when $RC_A$ reaches 0.

As discussed above, in SPS, UEs do not know the *RC* values of others, so some resource is unnecessarily marked unusable by its sensing component. For example, in Figure 2a, other UEs will sense the transmissions of *B* at subframe $n_B$ and *A* and $n_A$. Then they will consider the same single-subframe resources at $n_B + RRI$ and $n_A + RRI$ are not available. For those whose sensing windows contain these resource locations will have to vie for the remaining resource although the locations will be unused by *A* and *B*. When the channel busy ratio (CBR) [39] is high so that the available resources are limited, such inefficiency would be highly undesirable. In contrast, in SPS/LA, such inefficiency is prevented. Not because SPS/LA explicitly transmits the *RC* count, but because it broadcasts the planned initial resource location of the next streak. The *RC* count can be inferred to reach zero within *T* of the planned location. As with explicit RC advertisement [32], when the currently used location will become unused, which other UEs can use it in SPS/LA.

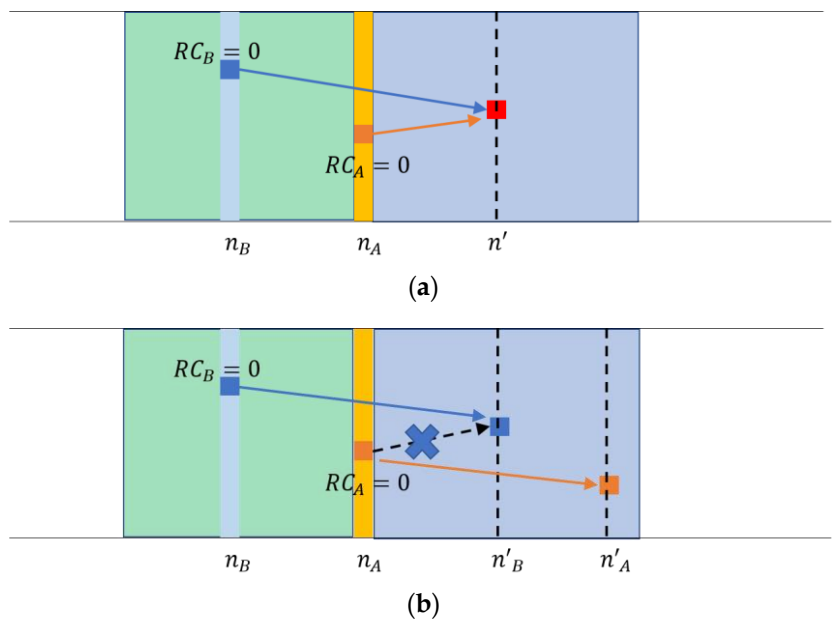

**Figure 2.** Use of the lookahead. (**a**) SPS cannot resolve the collision if two close UEs choose the same resource. (**b**) SPS/LA can resolve most potential collisions; *A* knows *B*'s planned movement through the received lookahead in *B*'s packet

Finally, we note that the use of the lookahead does not replace the sensing mechanism. As in the original SPS algorithm, the UE that reselects at $RC = 0$ should check the sensing window for the resource use history, in addition to the lookaheads that it has received that fall in the selection window. In essence, the lookahead is a reinforcement of the existing sensing-based SPS algorithm, not a replacement.

### 4.3. SPS/LA Algorithm

Algorithm 2 shows the SPS/LA logic. There are three more variables, all related to the lookahead: *LATxSubCh*, *LATxSubframe*, *LARC* (line 7). These are selected when $RC = RC_{LA}$ (lines 14, 16, 17, 19). In case we stay in the current resource location, the subchannel is maintained (line 17), but the first subframe index is set to the starting subframe for the first packet in the next streak (line 16). In case we need to move, the two values are selected using the lookaheads received from other UEs (line 19). When the current streak ends, the three values are used to set the RC counter (line 25) and the time-frequency coordinate (lines 27, 28, 30). However, before we commit, we double-check if the lookahead location is still available using both the lookaheads from other nodes and the sensed resource use map (line 26). In case the lookahead location is likely to be occupied, we revert to the SPS algorithm (line 30). Regarding this last case, Figure 3 shows an example. Suppose *A* and *B* are two vehicular UEs that happen to be transmitting their BSMs in the same subframes for their current streak. The current RC value of $A$ is one smaller than $B$'s. When $A$'s RC value reaches $RC_{LA}$, it plans for the next streak whose first packet will be transmitted at subframe $n^t$. In one RRI, $B$ also gets to choose its planned resource. Since these two are transmitting in the same subframes, however, they cannot hear each other and are not aware of each other's plans. If they happen to choose the same resource located at $n^t + RRI$, they will collide without knowing it, that is, without the double-check in line 30. The double-check performed at $n + RRI \times (RC_{LA} + 1)$ by $B$, however, reveals that an actual BSM transmission took place at $n^t$. This is exactly one RRI before the $B$'s resource planned for its next streak. Thus, when $B$ executes the SPS select, it will be able to avoid the collision at subframe $n^t + RRI$.

---

**Algorithm 2** Reselection in SPS/LA

---

1: **procedure** LA-SPS ($RRI, T1, T2, C1, C2, RC_{LA}, N_{subCH}, probResourceKeep$)
2:     $txSubCH \leftarrow random(1, N_{subCH})$              *1>* Initializations for $txSubCH, txSubframe, RC$
3:     $txSubframe \leftarrow random(1, RRI)$
4:     $RC \leftarrow random(C1, C2)$
5:     $subframe \leftarrow 0$                                        *1>* This is the current time
6:
7:     $LATxSubCh, LATxSubframe, LARC \leftarrow 0$         *1>* Initialize the lookahead-related parameters
8:     – – – – – – – – – – – – – – – – – – – – – – – – – – – – – – – – – –
9:     **while** *True* **do**
10:         **if** $subframe == txSubframe$ **then**                      *1>* It is time to transmit
11:             $txPacket(txSubCH)$                     *1>* Transmit packet on the specified subchannel(s)
12:             **if** $RC \mathrel{/}= 0$ **then**
13:                 **if** $RC == RC_{LA}$ **then**            *1>* Time for early reselection; do not commit just yet
14:                     $LARC \leftarrow random(C1, C2)$                  *1>* Set the length of the next streak
15:                     **if** $random(0, 1) < probResourceKeep$ **then**       *1>* Next streak will stay here
16:                         $LATxSubframe \leftarrow txSubframe + RRI \times (RC + 1)$
17:                         $LATxSubCh \leftarrow txSubCH$          *1>* Inherit the current resource coordinate
18:                     **else**                                  *1>* Must move from current location
19:                         call **select_ resource_lookahead( )**  *1>* Select resource using lookaheads+SPS
20:                     **end if**
21:                 **end if**
22:                 $txSubframe \leftarrow txSubframe + RRI$  *1>* Schedule next transmission in the current streak
23:                 $RC \leftarrow RC - 1$
24:             **else**                                            *1>* Time to switch to next streak
25:                 $RC \leftarrow LARC$                        *1>* Next streak becomes current streak
26:                 **if** check_lookahead( ) **then**       *1>* Double-check if lookahead is still available
27:                     $txSubframe \leftarrow LATxSubframe$
28:                     $txSubCH \leftarrow LATxSubCh$                       *1>* Commit to the lookahead
29:                 **else**                         *1>* Resource map changed; lookahead occupied
30:                     call **select_ resource_lookahead( )**     *1>* Select resource using lookaheads+SPS
31:                 **end if**
32:             **end if**
33:         **else**
34:             call **sensing_update( )**          *1>* Keep sensing and update resource use map; Section 3.1
35:         **end if**
36:         $subframe \leftarrow subframe + 1$                                    *1>* Push time
37:     **end while**
38: **end procedure**

---

Notice that the SPS/LA algorithm schedules the resource in first-come-first-serve (FCFS). The UE that first sends its lookahead takes priority, and those that internally decided for the same location as the lookahead should change their decision when they see the first advertised lookahead. Unless two UEs decide on the lookahead in the same subframe, the UE that computes the lookahead later cannot pick the resource at an already planned location by other UEs. Therefore, there is no special resolution procedure for the case multiple UEs plan for the same resource location. In the case that UEs in the same subframe (but different subchannels) plans for the same resource location, the current lookahead scheme cannot resolve the collision (Figure 4). One potential solution to this problem is a third node that can hear these two UEs tell in its packet about the imminent collision. However, the probability should be small, so we defer such a solution to a future work.

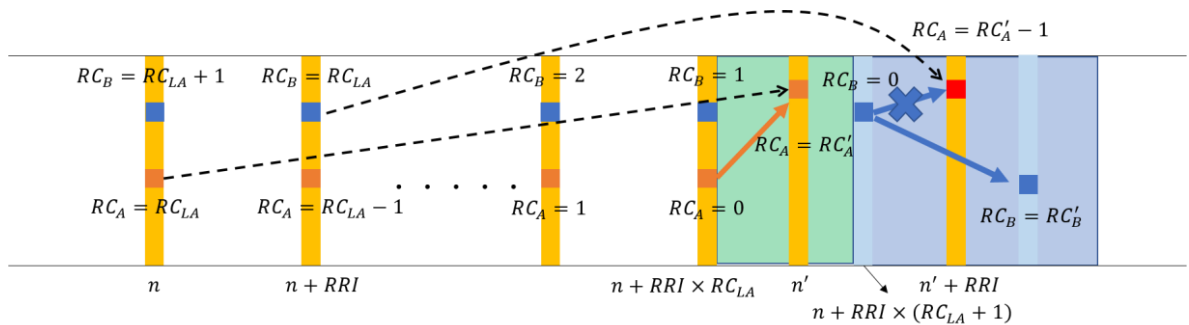

**Figure 3.** SPS sensing prevents the collision of the planned moves if RCs expire in different subframes.

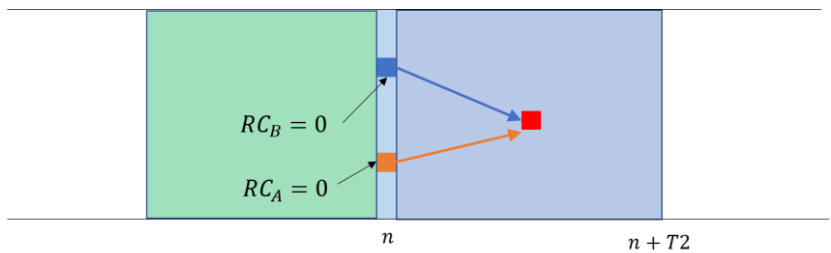

**Figure 4.** Collision is not detected by the UEs using the same subframe.

### 4.4. Implementation Considerations

In this paper, our focus is on exploring the performance impact of the lookahead scheme. Implementing it is not an immediate issue, but here we briefly mention possible approaches. First, we could include it in the BSM in Part II [2] where an optional information can be carried. However, it would pose difficulty because the SPS scheduling is far below the application layer in the protocol stack. We would need a cross-layer communication between the application layer that schedules BSM transmissions and the physical layer that controls the resource scheduling. Second, a more desirable way would be including the lookahead in the SCI Format 1 [38], which currently has the following information (numbers in parentheses are number of bits for each field).

· Priority (3)
· Resource reservation interval (RRI) (4)
· Frequency resource location of initial transmission and retransmission ($x = \lceil \log_2(N_{subCH}^{SL}(N_{subCH}^{SL} + 1)/2) \rceil$)
· Time gap between initial transmission and retransmission (4)
· Modulation and coding scheme (5)
· Retransmission index (1)
· Transmission format (1)
· Reserved information (14−$x$)

The additional data size to carry the lookahead is the time-frequency location of the transmission and the size of the resource allocation. One issue with this approach is that unless we increase the size of the SCI, there are not many remaining bits in the reserved information field to carry all four fields of a lookahead. Assuming that the lookahead information is added to the existing information above, we need $x (= \lceil \log_2(N_{subCH}^{SL}(N_{subCH}^{SL} + 1)/2) \rceil)$ bits to specify the frequency resource location $c$ and the length $L$ in the selected subframe. For example, if $N_{subCH}^{SL} = 25$, we need 9 bits to specify the two values. In this case, we could use the remaining bits in the reserved information field. Since the subframe specified in the lookahead is not the current subframe, we additionally need bits to encode the subframe number $n$ in the lookahead. As each vehicle transmits a beacon at least once a second, the offset to the subframe is within 1000 subframes of the current subframe at which the lookahead is being determined. To encode this offset, we need 10 bits. Again, we could use the remaining 6 bits from the added bytes to cope with larger $N_{subCH}^{SL}$. Figure 5 shows the additional number of bits that we would require by including the lookahead in the SCI, as a function of $N_{subCH}^{SL}$ and the offset of the subframe designated for the lookahead. The figure assumes that we do not borrow the bits from the reserved information field. If the overhead is excessive to offer in the SCI, we could even consider carrying it in the BSM as discussed in the first alternative or as in He et al. [31] at the cost of increased cross-layer signaling. However, we defer this implementation issue to a future work, and in this paper, we focus on the potential impacts of using such lookahead on the message collision mitigation performance.

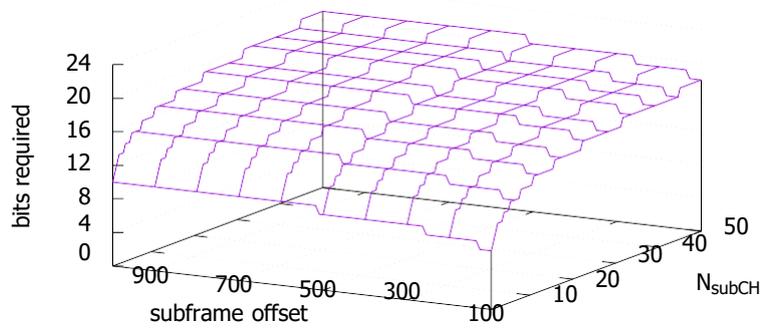

**Figure 5.** Number of additional bits to include the lookahead in the SCI Format 1, including the ten offset bits to the subframe in the lookahead.

## 5. Performance Evaluation

### 5.1. Analysis

Here, we perform an average-case analysis of SPS to understand its dynamics in terms of the many parameters it uses. In particular, we focus on the collision probability upon (re)selection because under the sensing and/or the lookahead, it is the only occasion when the collision can take place. Although the collision upon the (re)selection leads to a multiple packet collision because of the semi-persistency, the number of such ensued packet collisions will be identical for both schemes. Thus, we focus on the collision probability upon (re)selection.

To facilitate analysis, we apply some simplifying assumptions. First, we set $probResourceKeep = 0$, namely, we do not keep the current resource upon reselection [35]. Second, we assume that the vehicle population is static, namely no churn in the membership. Suppose the vehicular UE $V$ reached $RC = 0$ at subframe $n$. Let $N_{subCH}$ is the number of subchannels in the given carrier, and $T = T2 - T1 + 1$ is the selection window width. In the reselection for a vehicle $V$, there are $R_{tot} = N_{subCH} \cdot T$ single-subframe resources in the selection window. Among them, $R = R_{tot}(1 - CBR)$ are available for reselection. Here, CBR measured in subframe $n$ is defined to be the portion of subchannels in the resource pool whose RSSI measured by the UE exceed a (pre-)configured threshold sensed over subframes $[n-100, n-1]$ [39]. Essentially, it is the resource use. Although the UEs that reselect in $[n-T+1, n]$ will not use their current resource locations because $probResourceKeep = 0$, other UEs do not use these locations because of their sensing result in the previous RRI (see discussion in Section 4), so $R$ accounts for this inefficiency as well. In each subframe, there are $N_{sf} = N_{subCH} \cdot CBR$ UEs transmitting on average. In subframe $n$, there are $N_{sf}^{\mathsf{t}} = N_{sf}/C$ nodes on average whose RC counter reached 0 including $V$, where $C = (C1 + C2)/2$. For simplicity, we assume that every UE has the same RRI of 100 ms so that every UE can reselect only a single resource in the selection window. Any one of these contending UEs with $RC = 0$ reselects a resource location in $R$ with probability $p = 1/R$. In subframe $n$, the probability $P_{col,n}$ that a UE $V$ collides with other UE(s) on the same subframe $n$ is given by

$$P_{col,n} = 1 - (1 - p)^{N_{sf}^t - 1} \quad (1)$$

In subframe $n - k$ $(0 < k \leq T - 1)$, there are $N_{sf}^t$ UEs that expired their RCs that can be contending with $V$. With $V$, they share $T - k$ subframes in the selection window. The probability that a UE there selects one resource in the overlapped window is $p^{(n-k)} = (T - k)/T \cdot p$. As above, the probability $P_{col,n-k}$ that a UE $V$ collides with other UE(s) on this subframe $n - k$ is given by

$$P_{col,n-k} = 1 - \left\{ 1 - p^{(n-k)} \right\}^{N_{sf}^t} \quad (2)$$

Then, the total collision probability for $V$ can be computed as

$$
\begin{aligned}
P_{col} &= 1 - \prod_{k=0}^{T-1} (1 - P_{col,n-k}) \\
&\approx 1 - \prod_{k=0}^{T-1} \left( 1 - p^{(n-k)} \right)^{N_{sf}^t} \\
&= 1 - \prod_{k=0}^{T-1} \left( 1 - \frac{T-k}{T} \times \frac{1}{N_{subCH} \cdot T \cdot (1-CBR)} \right)^{N_{subCH} \cdot CBR/C}
\end{aligned}
\quad (3)
$$

Notice that $P_{col}$ is a function of the system parameters $T1$, $T2$, $N_{subCH}$, $C1$, $C2$, and the traffic load (CBR).

In contrast, the SPS/LA scheme removes the collisions caused by the UEs transmitting in the past subframes (Equation (2)), hence $P_{col} = P_{col,n}$. This is because these UEs make their plans known, $V$ can avoid the locations they plan to use in its sensing window. The only collisions are caused by the UEs in the same subframe as $V$ with their RC expiring, whose transmissions are mutually hidden (Figure 4). Figure 6 shows the collision probabilities upon (re)selection for the two schemes as a function of CBR. We observe that the collision probability is more than an order of magnitude lower in SPS/LA. In practice, the collision probability of the SPS/LA grows higher than Equation (1). It is due to the reverting to the original SPS selection that we need for the last-minute check before the final commit to the lookahead (lines 26 and 30 in Algorithm 2). Especially when the CBR increase to a high number, such reversions are more likely. Nevertheless, Figure 6 tells us how an additional piece of information regarding each UE's plan of resource use can reduce message collisions in SPS. We will confirm it through simulations below, under more various settings.

## 5.2. Simulation

In this section, we conduct simulations to verify the analysis above. Since we focus on the purely algorithmic aspect of the sensing-based SPS and our proposal to enhance it, and the problem formulation is simple, we use a homegrown simulator written in Python. We run ten instances of each simulation instance to obtain narrow enough confidence intervals for the results. We assume that UEs are static, and for the channel propagation, we assume the simple disc model where all UEs within the communication range enjoy the perfect physical channel. The only exception to the static UE locations is one experiment where the population churn, due to new UEs joining in the communication range and others leaving, is modeled and its impact is observed. Please note that our intention in this paper is to demonstrate the point that even under perfect physical-layer conditions, the current standard SPS algorithm exhibits serious shortcomings in resolving resource conflicts and providing high-reliability communication. Due to the perfect propagation model, the spatial distribution of the UEs is immaterial. Towards the end of this paper, however, we briefly discuss the expected impacts of a more realistic physical configurations in terms of the wireless channel. In particular, we consider how such imperfect physical channel will affect the use of the lookahead and the consequent performance impacts.

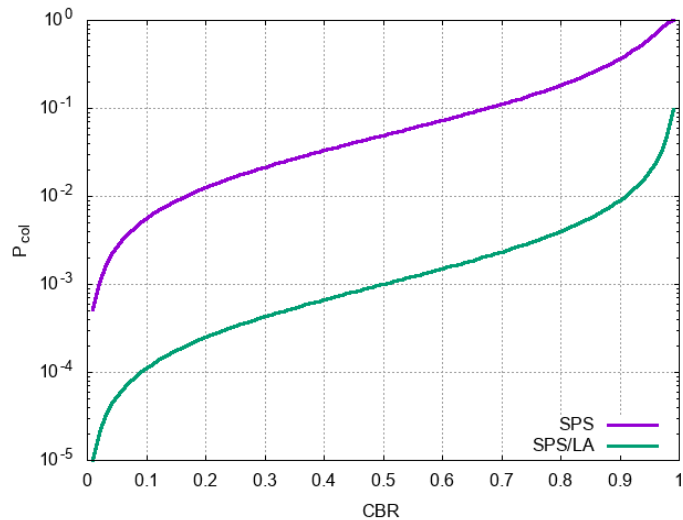

**Figure 6.** Collision probabilities upon reselection under SPS and SPS/LA.

For the simulation experiments, we use the following assumptions. First, the packet size is uniform at 300 bytes, and each packet requires a single-subframe resource. (For the optimization focusing on the uneven packet sizes, readers are referred to Molina-Masegosa [33].) Second, the sensing-based candidate resource selections (i.e., $S_A$ and $S_B$) have been already performed, at which instant our simulation starts. Packets are periodically transmitted every $RRI = 100$ ms. Following the standard, we set the RC range at $(C1, C2) = (5, 15)$. For the selection window, we set $(T1, T2) = (1, 100)$. We turn off the retransmission, so each BSM is transmitted only once. For the resource pool, we assume that a 20-MHz channel at 5.9 GHz is completely dedicated to Mode 4 using adjacent PSCCH + PSSCH subchannelization. We assume that there are $N_{subCH} = 25$ subchannels of 8 RBs each are defined per subframe. The maximum number of UEs is 2500 if their RRI is 100 ms, and 500 if RRI is 20 ms. The simulation parameters are summarized in Table 1.

**Table 1.** Parameters used in simulation.

| Parameter | Default Value |
|---|---|
| Retransmission | off |
| RRI | 100 ms |
| Sensing window | 1000 ms |
| BSM size | 300 bytes |
| $RC_{LA}$ | 1 |
| $L_{subCH}$ | 1 |
| $N_{subCH}$ | 25 |
| $RB/subCH$ | 8 |
| $T1, T2$ | 1, 100 |
| $C1, C2$ | 5, 15 |
| $probResourceKeep$ | 0 |

For the performance metric, here we use the collision probability. Please note that it is not the collision probability of the reselection operation that we considered in the analysis above. Instead, we use the fraction of the single-frame resources to which more than one packet arrived, out of the entire set of single-subframe resources generated throughout the simulation. Each simulation lasts 100 s, so the total number of single-subframe resources are 2,500,000. The presented results are averages of 10 simulation runs.

### 5.2.1. No Population Churn

In the first set of experiments, we consider the case where we have a stable topology, so there are no new vehicles joining or existing members leaving the communication range. We set the probability *probResourceKeep* = 0, so that when RC reaches 0, a UE always reselects a different resource. We divide the CBR into two regimes: 1–5% and 10–90%. To create the CBR from 1% to 5%, we place 25 to 125 vehicles in mutual communication range. To create the CBR from 10% to 90%, we place 250 to 2250 vehicles in mutual communication range.

Figure 7 compares the collision probabilities of SPS and SPS/LA. The collision probability plotted in the figure is computed as the total number of collisions divided by the total number of single-subframe resources during the elapsed time. For example, the value at 40 s is obtained by dividing the number of collisions up to that time by the number of single-subframe resources during that time ($=25 \times 40 \times 10^3 = 10^6$). Notice that the $y$-axis is in log scale. Under very light loads, the standard SPS scheme keeps the collision probability under $10^{-4}$. However, in the SPS/LA, the probability is even lower, orders of magnitudes lower (compare the $y$-axis with SPS). The collision probabilities approach $10^{-6}$. In particular, there was no collision in 1 and 2% CBR during 100 s of simulated time. It bears out our analysis where the reselection collisions are similarly reduced in SPS/LA.

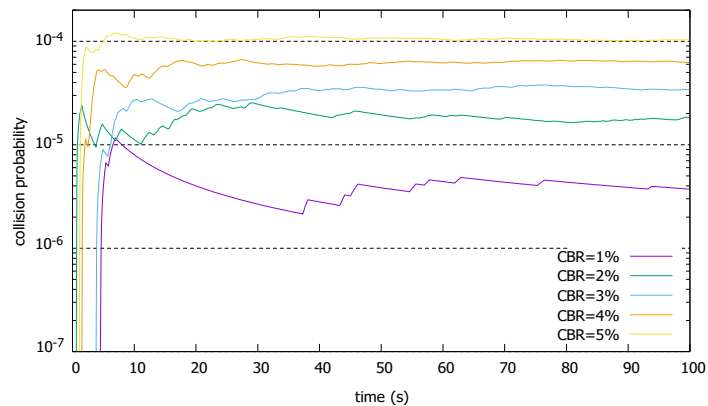

(a) SPS

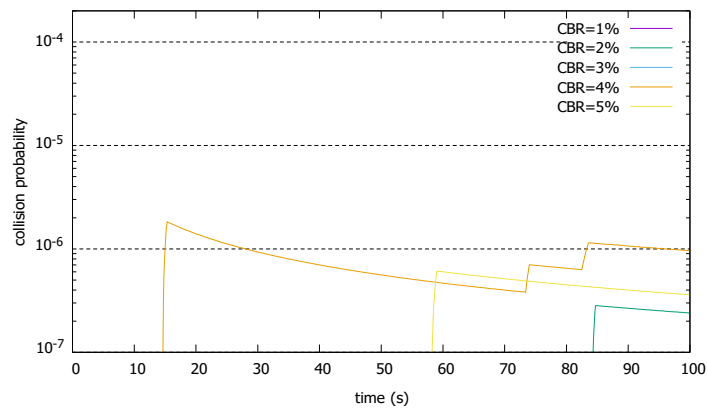

(b) SPS/LA

**Figure 7.** Packet collision rates comparison in (**a**) SPS and (**b**) SPS/LA, light load.

As we run ten instances of simulation for each data point, the confidence interval is very narrow, and hard to see with the log scale plots. Table 2 shows the 95% confidence interval at simulation time $t = 100$.

**Table 2.** Collision probability at $t = 100$ with 95% confidence interval (CI) width.

| CBR | Mean$_{SPS}$ | $|$CI$|_{SPS}$ | Mean$_{SPS/LA}$ | $|$CI$|_{SPS/LA}$ |
|---|---|---|---|---|
| 1% | $3.72 \times 10^{-6}$ | $6.33 \times 10^{-6}$ | 0 | 0 |
| 2% | $1.85 \times 10^{-5}$ | $1.38 \times 10^{-5}$ | $2.40 \times 10^{-7}$ | $1.09 \times 10^{-6}$ |
| 3% | $3.41 \times 10^{-5}$ | $1.34 \times 10^{-5}$ | 0 | 0 |
| 4% | $6.21 \times 10^{-5}$ | $2.31 \times 10^{-5}$ | $3.13 \times 10^{-6}$ | $4.34 \times 10^{-6}$ |
| 5% | $1.04 \times 10^{-4}$ | $1.96 \times 10^{-5}$ | $1.17 \times 10^{-6}$ | $1.63 \times 10^{-6}$ |

Now, we increase the CBR. Figure 8 shows the results. We notice that the collision probability far exceeds $10^{-4}$ in SPS even for 10% CBR. This number is important because some prominent applications envisioned for 5G eV2X require 99.99% reliability (Table 3) [40]. 5G eV2X will probably use a different radio resource format, but still the SPS performance is disappointing. When the CBR reaches 90%, the collision probability is 12.8%. In SPS/LA, we notice that the collision probability is reduced by a factor of 10 or more compared with the original SPS. Again, this result is in line with the reselection collision probability analysis. In particular, the SPS/LA scheme keeps the probability under $10^{-4}$ until the CBR exceeds 40%. Even when the CBR increase further, the performance gap between the two algorithms is maintained at more than an order of magnitude. For an extremely high CBR of 90%, the collision probability under SPS/LA is still under 0.6%, compared with 12.8% under SPS.

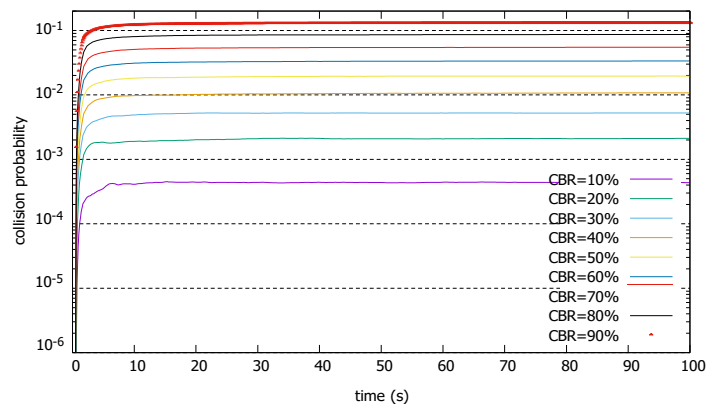

(a) SPS

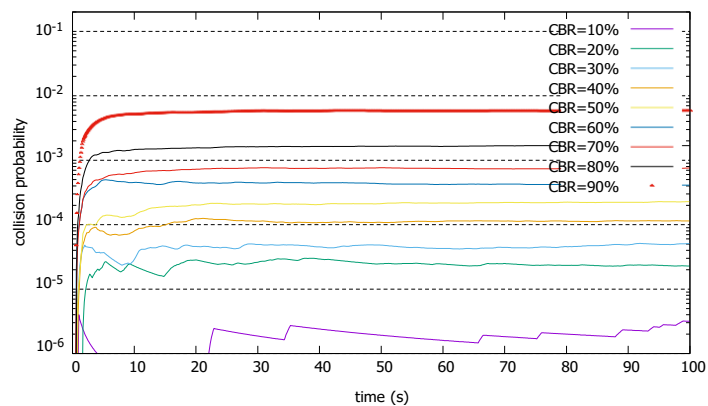

(b) SPS/LA

**Figure 8.** Packet collision rates comparison in (**a**) SPS and (**b**) SPS/LA, heavy load.

Again, we show the mean and the 95% confidence interval widths for each case in Table 4. We again see that the confidence interval width is very small. We obtain the same narrow confidence intervals, so for subsequent results, we omit them.

**Table 3.** Payload size, reliability, and latency requirements for prospective 5G eV2X applications.

| Application | Packet Size (B) | Reliability (%) | Latency (ms) |
|---|---|---|---|
| Vehicle Platooning | 300–400 | 90 | 25 |
| Remote Driving | 300–400 | 99.99 | 5 |
| Autonomous Cooperative Driving | 1200 | 99.99 | 10 |
| Collective Perception of Environment | 1600 | 99 | 100 |
| Cooperative Collision Avoidance | 2000 | 99.99 | 10 |

**Table 4.** Collision probability at $t = 100$ with 95% confidence interval (CI) width.

| CBR | Mean$_{SPS}$ | $|$CI$|_{SPS}$ | Mean$_{SPS/LA}$ | $|$CI$|_{SPS/LA}$ |
|---|---|---|---|---|
| 10% | $4.34 \times 10^{-4}$ | $4.18 \times 10^{-5}$ | $3.16 \times 10^{-6}$ | $6.16 \times 10^{-6}$ |
| 20% | $2.11 \times 10^{-3}$ | $1.22 \times 10^{-4}$ | $2.27 \times 10^{-5}$ | $1.20 \times 10^{-5}$ |
| 30% | $5.24 \times 10^{-3}$ | $2.83 \times 10^{-4}$ | $5.08 \times 10^{-5}$ | $1.67 \times 10^{-5}$ |
| 40% | $1.08 \times 10^{-2}$ | $3.05 \times 10^{-4}$ | $1.15 \times 10^{-4}$ | $1.26 \times 10^{-5}$ |
| 50% | $1.96 \times 10^{-2}$ | $2.73 \times 10^{-4}$ | $2.27 \times 10^{-4}$ | $4.45 \times 10^{-5}$ |
| 60% | $3.36 \times 10^{-2}$ | $3.72 \times 10^{-4}$ | $4.10 \times 10^{-4}$ | $5.85 \times 10^{-5}$ |
| 70% | $5.49 \times 10^{-2}$ | $3.64 \times 10^{-4}$ | $7.56 \times 10^{-4}$ | $7.58 \times 10^{-5}$ |
| 80% | $8.65 \times 10^{-2}$ | $6.08 \times 10^{-4}$ | $1.69 \times 10^{-3}$ | $1.30 \times 10^{-4}$ |
| 90% | $1.28 \times 10^{-1}$ | $5.06 \times 10^{-4}$ | $5.76 \times 10^{-3}$ | $3.20 \times 10^{-4}$ |

Since we did not allow any population churn in this experiment, the only collisions arise from the reselection at $RC = 0$ with the blind choice of the resource for the next streak. The lookahead turns the blinded ventures into the enlightened choices in SPS/LA, thereby significantly reducing the collision probability.

### 5.2.2. Nonzero Resource Keeping Probability

Here, we again assume that there is no population churn. It is to focus on the impact of the moving probability *probResourceKeep*. Figure 9 compares the collision probabilities of SPS and SPS/LA under various *probResourceKeep*, ranging from 0.2 to 0.8. To give the readers more direct sense of the performance differences, henceforth we use the linear scale on the *y*-axis. We present only the collision probability at the end of the simulation when the probability has been stabilized.

First, we notice that there is impact from varying *probResourceKeep* in SPS. Specifically, as the resource keeping probability increases, the collision probability decreases. This result is in line with other works [30]. Also, it confirms our earlier observation that the collisions are a direct consequence of blind reselection. Nevertheless, we cannot simply prohibit reselection to reduce the collision probability because the reselection also serves to accommodate topology dynamics in the vehicle traffic. In Section 5.2.3, we will observe the impact of *probResourceKeep* under more dynamic population changes.

Compared to SPS, the SPS/LA scheme rarely produces collisions under the light load. Specifically, the collision probability, if not zero, is frequently orders of magnitudes smaller than in SPS.

Under heavier traffic loads in Figure 10, we first notice that the collision probability superlinearly increases in CBR in both schemes. Here again, increasing *probResourceKeep* has a collision mitigating effect. We can also clearly observe that SPS/LA significantly reduces the collision probability, maintaining the order-of-magnitude performance gap.

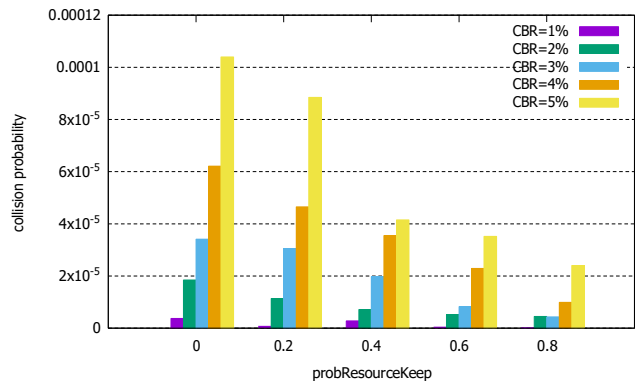

(a) SPS

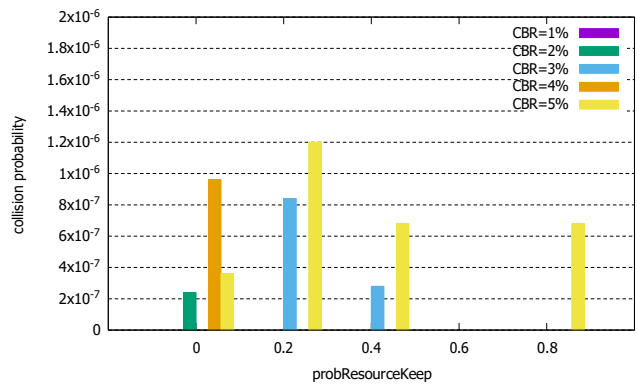

(b) SPS/LA

**Figure 9.** Collision probabilities under varying *probResourceKeep*, light load. (**a**) SPS. (**b**) SPS/LA.

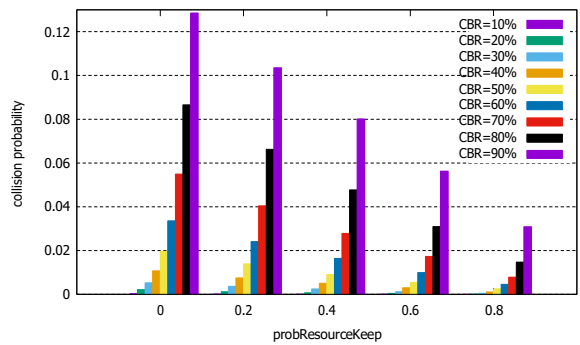

(a) SPS

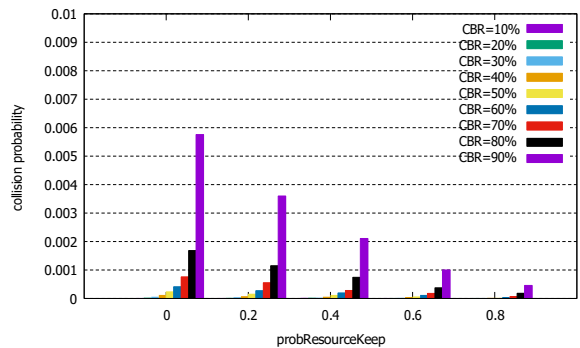

(b) SPS/LA

**Figure 10.** Collision probabilities under varying *probResourceKeep*, heavy load. (**a**) SPS. (**b**) SPS/LA.

### 5.2.3. Population Churn

Now, we allow the population churn. This is necessary for more realistic scenario because vehicles come into and go out of the communication range even in stable driving environments. It is even more dynamic when roads merge or in the intersection, etc.. To simply the experiment setting, we let new vehicles come in to the communicating group of vehicles at a churn rate $\lambda$ (%/s) and some in the group exit at the same rate. For example, the churn rate of $\lambda = 0.01$ at $CBR = 20\%$ corresponds to 5 vehicles newly joining/leaving from the entire population of 500 vehicles every second. It simulates a very stable group of vehicles in terms of churn. On the other hand, $\lambda = 0.2$ at $CBR = 90\%$ corresponds to 450 vehicles changing per second, which is a very high dynamics scenario. We assume that the newly joining UEs randomly choose their initial resource locations, without any chance for sensing. Please note that it is a rather harsh assumption because as vehicles join, they will be able to perform sensing at least for a short duration. However, to stress the two schemes, we assume so.

Figure 11 shows the results. We observe that with the churn rate increasing, the collision probability increases in both schemes. This is in line with the observation made by Bonjorn et al. [32]. It is expected because under our harsh assumption above, the initial collisions are not preventable. We also observe that SPS/LA is more heavily affected by the harsh assumption. It reflects the fact that the SPS/LA scheme effectively controls the other type of collisions, namely those arising from the lack of coordination among vehicles as to the resource reselection in SPS. In contrast, the SPS performance only slightly degraded. It implies that for the SPS without the lookahead, most of the collisions are of similar blind nature to the newly joining vehicles in this experiment. Overall, the SPS/LA performs better than SPS, even with very high churn rates.

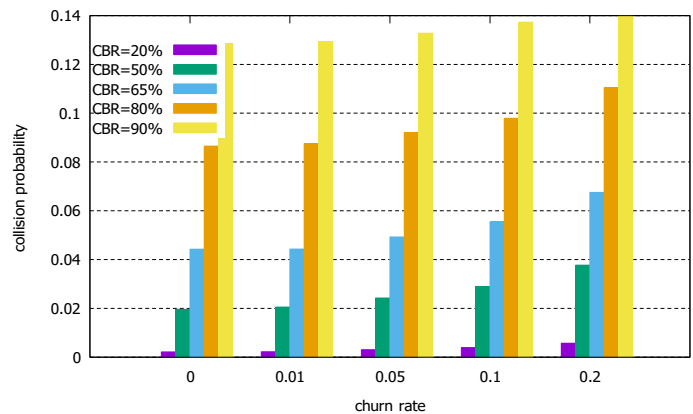

(a) SPS

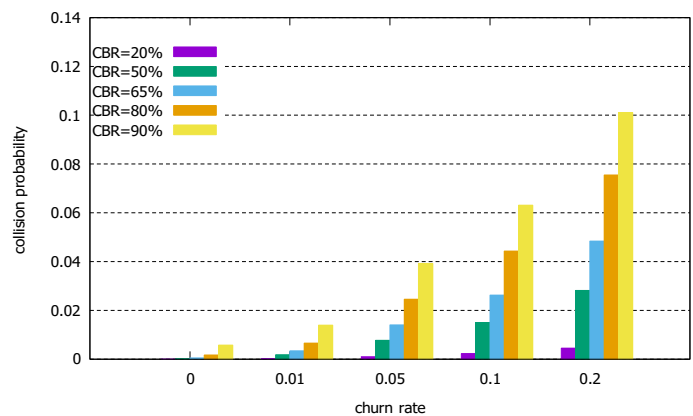

(b) SPS/LA

**Figure 11.** Collision probabilities under population churn. (**a**) SPS. (**b**) SPS/LA.

*5.3. Channel Impacts on the Proposed Algorithm*

In this paper, we focused on the algorithmic aspects of the sensing-based SPS algorithm, and abstract away the physical-layer issues such as coverage, path loss, fading, and hidden terminals. We intended to demonstrate the point that even under perfect physical-layer conditions, the current standard SPS algorithm exhibits serious shortcomings in resolving resource conflicts and providing high-reliability communication. However, to evaluate the performance impact of the proposed lookahead scheme more precisely, we would include the physical-layer properties in the simulation. We defer it to a future work, but here we provide a brief discussion on the expected impacts of the physical-layer features.

In case there are hidden terminals attempting to reserve an overlapping set of resources using the lookahead, the UEs that are in the hearing distance of both the hidden terminals will not be able to hear from either. However, this problem is not specific to our proposed scheme. It is also a problem for the sensing-based SPS as well. In fact, it is a problem for any resource reservation scheme based on the physical sensing. Even the virtual carrier sensing feature in 802.11 is not applicable in DSRC because the beacon messages are broadcast. Therefore, we believe that this problem is beyond the scope of this paper and should be separately addressed.

In case the lookahead is not received correctly due to various physical-layer degenerations such as path loss and fading, the receiving UEs can consider the reserved resource as still available. The lack of lookahead information, however, reverts the system back to the original sensing-based SPS algorithm because in our algorithm all UEs will check the validity of their reservations when their RC reaches zero, regardless of their published lookahead (line 26, Algorithm 2). Therefore, the addition of lookahead does not produce a side-effect when combined with its loss due to the adverse channel conditions.

Finally, let us consider the case where two UEs, A and B, have the asymmetric channel condition. Without loss of generality, let us suppose A can hear the reservation made by B through the lookahead but not vice versa. In our algorithm, the node that sees a resource block being reserved by other UE should change its target (Section 4.3). Thus, if only A hears B making the reservation first, A will choose another resource as the algorithm intends. In case A makes the first reservation through the lookahead but B cannot hear it, B will go ahead and make the reservation for the overlapping resource through the lookahead. If A hears it before it commits at $RC = 0$, A recognizes the asymmetric channel condition and should change its reservation. If B does not make the reservation before A finally commits at $RC = 0$, the case becomes similar to Figure 4, except that A and B do not have to use the same subframe to experience the problem. However, again, this problem does not make the proposed algorithm worse than the original SPS algorithm. In the SPS, A and B do not even have a chance to realize the conflict at all. The lookahead scheme at least has a chance to resolve one case where A can hear B.

## 6. Conclusions

By drastically reducing the uncertainties in the choice of the resource to use for the next string of messages, we showed that we can significantly reduce the message collisions in the C-V2X sidelink Mode 4. We proposed the use of the "lookahead", that contains the next starting resource location in the time-frequency plane. By exchanging the lookahead information piggybacked on the periodic safety message, we can eliminate most message collisions arising from the ignorance of other UE's internal decisions. The only required cost to implement our lookahead scheme is the inclusion of the lookahead in the control part of the packet. Currently, the SCI Format 1 for Mode 4 communication is too small to accommodate the three numbers that describe the location and the size of the resource planned for the next semi-persistent burst of packets. Considering the stringent reliability requirement in future C-V2X applications such as autonomous driving, we may well weigh the possibility of including the lookahead information in the next version of the SCI to improve the SPS performance in the next generation of C-V2X communication.

**Author Contributions:** H.K. conceived and designed the experiments; Y.J. and S.K. performed the experiments and analyzed the data; H.K. wrote the paper.

**Funding:** This research was supported by a grant (18CTAP-C133064-02) from Technology Advancement Research Program (TARP) funded by Ministry of Land, Infrastructure and Transport of Korean government.

**Conflicts of Interest:** The authors declare no conflict of interest.